\documentclass[sigconf]{acmart}
\AtBeginDocument{%
  \providecommand\BibTeX{{%
    \normalfont B\kern-0.5em{\scshape i\kern-0.25em b}\kern-0.8em\TeX}}}

\copyrightyear{2023} 
\acmYear{2023} 
\setcopyright{acmlicensed}
\acmConference[IDC '23]{Interaction Design and Children}{June 19--23, 2023}{Chicago, IL, USA}
\acmBooktitle{Interaction Design and Children (IDC '23), June 19--23, 2023, Chicago, IL, USA}
\acmPrice{15.00}
\acmDOI{10.1145/3585088.3589388}
\acmISBN{979-8-4007-0131-3/23/06}

\begin{document}




\title[Exploring Children's Preferences for a Homework Companion Robot]{\textit{``My Unconditional Homework Buddy:''} Exploring Children's Preferences for a Homework Companion Robot}



\author{Bengisu Cagiltay}
\email{bengisu@cs.wisc.edu}
\affiliation{%
  \institution{Computer Sciences Department, University of Wisconsin-Madison}
  \city{Madison}
  \state{WI}
  \country{USA}
}

\author{Bilge Mutlu}
\email{bilge@cs.wisc.edu}
\affiliation{%
  \institution{Computer Sciences Department, University of Wisconsin-Madison}
  \city{Madison}
  \state{WI}
  \country{USA}
}

\author{Joseph E Michaelis}
\email{jmich@uic.edu}
\affiliation{%
  \institution{Learning Sciences Research Institute, University of Illinois Chicago}
  \city{Chicago}
  \state{IL}
  \country{USA}
}

\renewcommand{\shortauthors}{Bengisu Cagiltay, Bilge Mutlu, and Joseph Michaelis}



\begin{abstract}
We aim to design robotic educational support systems that can promote socially and intellectually meaningful learning experiences for students while they complete school work outside of class. To pursue this goal, we conducted participatory design studies with 10 children (aged 10--12) to explore their design needs for robot-assisted homework. We investigated children's current ways of doing homework, the type of support they receive while doing homework, and co-designed the speech and expressiveness of a homework companion robot. Children and parents attending our design sessions explained that an emotionally expressive social robot as a homework aid can support students' motivation and engagement, as well as their affective state. Children primarily perceived the robot as a dedicated assistant at home, capable of forming meaningful friendships, or a shared classroom learning resource.
We present key design recommendations to support students' homework experiences with a learning companion robot.
\end{abstract}




\begin{teaserfigure}
  \includegraphics[width=\textwidth]{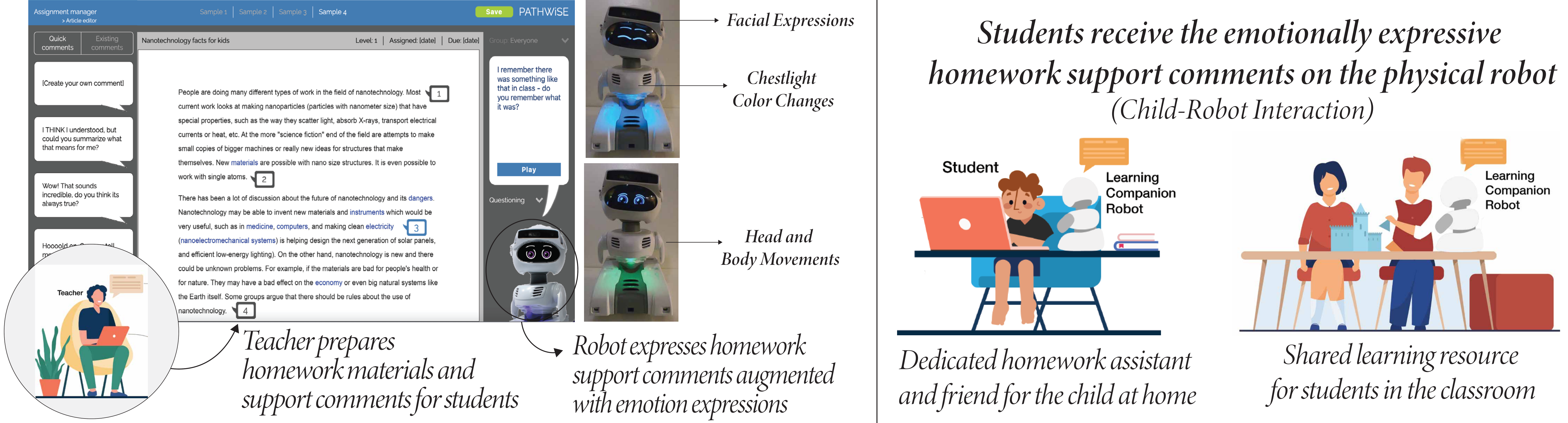}
  \caption{We explored students' preferences for augmented homework experiences with a \textit{homework companion robot.}}
  \Description{This figure has two main panels. The panel on the left shows a screenshot of a homework tool, an icon of a teacher, and text stating that "teacher prepares homework materials and support comments for students." In addition, this panel includes three photos of a social robot Misty, that is expressing different emotions. Labels point to the robot's facial expressions, chestlight and color changes, head and body movements. The text states that "robot expresses homework support comments augmented with emotion expressions. On the right panel, the main title states "students receive the emotionally expressive homework support comments on the physical robot (child-robot interaction). Below the title, there are two illustrations: one that shows a student interacting with a social robot on their own with a text stating "dedicated homework assistant and friend for the child at home", and second shows two students interacting with the social robot together with a text stating "shared learning resource for students in the classroom."}
  \label{fig:teaser}
\end{teaserfigure}

\maketitle

\section{Introduction}
Students need learning environments that encourage socially and intellectually meaningful interactions to support their motivation and learning experiences, particularly in STEM education~\cite{koenka_personalized_2019, hamilton_toward_2010}. Social interactions between teachers and students, rather than simply completing activities or exhibiting mastery on standardized tests, can support students' long-term knowledge building and deeper learning~\cite{bereiter_education_2002}. Teachers frequently deliver guidance to students through socially engaging interactions in classrooms, but this type of guidance would also benefit students outside of class. However, teachers struggle to extend in-class guidance to at-home activities, and parents often find it difficult to provide homework assistance to students~\cite{van_voorhis_costs_2011, hatch_making_2022}.

Robots, like other computerized learning systems, can quickly and effectively interact with students to provide additional support and guidance with homework. 
Several studies~\cite{bainbridge_benefits_2011, belpaeme_social_2018, li_benefit_2015, han2008comparative} have demonstrated that in educational contexts, social robots have a greater capacity for social interaction with children than computer-based digital agents, making them a suitable tool for social learning. Social robots designed to interact with students in a natural and intuitive way can provide support while they complete their homework assignments.
Humans tend to attribute human-like characteristics to nonhuman entities~\cite{epley_seeing_2007, kahn_robovie_2012, breazeal_emotion_2003, nass_machines_2000} and build social connections, such as empathy and rapport, with robots~\cite{darling_empathic_2015, darling_whos_2015} due to their anthropomorphic features or other design elements like eye gaze patterns~\cite{andrist_conversational_2014, mutlu_designing_2011} and references to previous interactions or personal backstories~\cite{kanda_two-month_2007, leite_social_2013, lee2022unboxing}. In special education, robots can support social communication and skills such as joint attention of children with autism~\cite{amanatiadis2020social} and in informal learning settings robots can promote learning and interest of children \cite{ho2021robomath, michaelis_reading_2018, michaelis_supporting_2019}. 

In an effort to improve the learning experience for students, we propose the use of social robots as learning companions that can transform homework experiences to promote social and intellectually meaningful learning. Our long-term objective is to design robotic educational support systems that provide students with socially and intellectually meaningful learning experiences.
In this paper, we explore the design of social learning companion robots to support students' homework experiences.
We ask the following research question:
\textit{\textbf{RQ: What needs do students have for interactions with a socially situated homework companion robot?}}
The contribution of this work is three fold. First, through participatory design sessions, we report students' perceptions and needs for receiving socially situated guidance in robot-assisted homework. Second, we co-created a preliminary design for student-robot interactions for robot assisted homework. Third, the outcome of this research builds theory in child-robot interaction and provides practical design guidance on student needs for augmented homework experiences and interactions with a homework companion robot.

\section{Related Work}

\subsection{Social Learning}
Learning sciences research has shown that social interactions are crucial for fostering deep knowledge and long-term interest in science~\cite{greeno_early_2016, vygotsky_mind_1978}. Deep knowledge is supported by embracing learning as a developmental process within a socio-cultural context~\cite{greeno_learning_2014} that benefits from interactions with people and environmental artifacts like computers and robots~\cite{hollan_distributed_2000, puntambekar_scaffolding_2009, sawyer_new_2014}. Social interactions with others can support comprehension, knowledge construction, and the synthesis of new ideas. In classroom interactions, teachers and peers can direct each other's attention during shared activities through discussion and gesture~\cite{fountas_guided_2012}.
Social interactions give the learner ways to build relationships and share common values with others~\cite{basu_developing_2007}, can support relationships, values, meaning~\cite{nieswandt_undergraduate_2015,palmer_student_2009}, and connections to foster interest and identity building~\cite{bergin_social_2016, krapp_basic_2005}. Social others can encourage students' interest in science through participating in learning activities and emphasizing the significance of science to students and their community~\cite{sansone_interest_2005}.

Regularly completing homework is associated with higher academic performance of students, but students who have difficulty with it could experience struggles in academic achievement. \cite{cooper_battle_2015}. 
Studies have shown that working with others on homework is a more positive experience for middle school students~\cite{kackar_age_2011}, and that guidance from others is a major factor in academic success~\cite{ramirez_low-income_2014}. However, children often receive little or no parent support or peer involvement in their homework, parents find helping with homework challenging due to the large time commitment and limited communication from teachers about the structure and meaning of the work~\cite{van_voorhis_costs_2011}. \citet{van_voorhis_costs_2011} found that when teachers consciously involve parents in students homework, students and families reported more positive emotions and attitudes towards the homework experience. Students need direction during homework so that they can link concepts, find value in the homework, and deal with distractions or drops in motivation~\cite{xu_why_2013}.
%

\subsection{Social Robots as Learning Partners} 
 Research has shown that social robots are more effective at promoting engagement and learning than virtual agents on computers~\cite{belpaeme_social_2018}. 
Children expect in-home social robots to take the role of a companion or assistant~\cite{cagiltay2020investigating} and hope for social robots to have valuable knowledge that could be realized in the form of intellectual assistance in tasks such as supporting homework~\cite{rubegni2022don}. The different roles that robots take as learning partners, such as tutee, tutor, or a peer agent, can impact children's learning and engagement~\cite{chen2020teaching}. While a robot \textit{tutor} can support children's learning, children exhibit greater affective states with a robot \textit{tutee}, and children's learning and engagement is mostly supported by a \textit{peer} robot.
Similarly, learning partner robots with different instructional styles (e.g., lecture, cooperative, and self-directed) and different speech styles (e.g., human-like and robot-like) could impact the learning experiences of children~\cite{okita2009learning}. 
Furthermore, children can recognize and internalize a growth mindset expressed by a peer-like social robot~\cite{park2017growing}.

Robots as learning partners should also have the ability to sustain motivation and engagement of students in the long-term, particularly in classroom or in-home settings \cite{kanda2012children, Cagiltay_engagement_2022}, where robots would be deployed for learning sessions over certain period of time. For example, \citet{kanda2012children} found that student motivation and encouragement can be supported through robot behaviors tailored to facilitate social connections and relationships, but these effects diminished over time ~\cite{kanda2012children}. 
In unsupervised naturalistic interactions with an in-home reading companion robot, children were often observed to include different family members such as siblings, parents, or grandparents in their reading interaction and share anecdotal moments with other family members which supported the maintenance of long-term engagement with the robot~\cite{Cagiltay_engagement_2022, offscript_HRI}.
In an analysis of primary school students interactions during a collaborative inquiry learning assignment, \citet{davison2020designing} identified 14 design goals for social robots that support children's learning. Three of these design goals included the robot's display of emotional responses, offering emotional support to the student, and engaging social in activities beyond learning. 
In sum, social robots as learning partners should support building and maintaining a positive social relationship with the child~\cite{davison2020designing}.

\subsection{Interaction Design with and for Children}
To ensure educational technologies and tools are designed with teachers' and students' needs and perspectives in mind, it is important to take a learner-centered design approach that would actively engage learners, teachers, and other community members as design partners to allow for an inclusive and iterative design process~\cite{disalvo2017participatory}.
A growing number of educational technology studies employ participatory design to build partnership between educators, students, parents, designers, developers~\cite{carroll2002building, borges2016participatory} ranging from designing educational and tangible robots~\cite{bertel2013robots, winkle2021leador, pnevmatikos2022designing, axelsson2019participatory, kirlangicc2020storytelling}, technology designed for children with disabilities and special needs~\cite{pires2020exploring, pires2022learning, koushik2019storyblocks}, educational games~\cite{danielsson2006participatory, ismail2017pdedugame}, online safety and cyberbullying trainings for teens~\cite{cyberbullying,bowler2015cyberbullying}.
Furthermore, \cite{ceha2022identifying} explored teacher's preferences for in-class social robots, and identified different behaviors for teacher-robot and student-robot interactions. Teachers envisioned students to interact with the robot to receive guidance, emotional support, build connections through shared knowledge and \textit{``model desirable student behaviour, engage students through conversation, and make the material relatable not only with what it says, but how it says it.''}

Our research builds on teacher insights from our prior work \cite{pathwise}, which similarly examined teacher's preferences for an authoring tool for robot-assisted homework assignments. Teachers used a social robot-augmented homework studio to create homework interactions that resembled classroom interactions. Using this tool, teachers wrote socially supportive learning prompts, added emotion expressions, and tested them on a social robot. Teachers initially wrote formal, test-like prompts and questions, however after using the robot, teachers adjusted prompts to make more personal connections with students. Findings from \cite{pathwise} show that teachers focused more on the homework support comments' tone than the robot's emotions. We use these teacher insights as an iterative step to designing a homework companion robot with and for children.



\begin{table*}
    \caption{\textit{Participant Information ---} Family demographics and background.}
    \label{dem-table}
    \centering
    \small
    \begin{tabular}{p{0.13\textwidth}p{0.1\textwidth}p{0.1\textwidth}p{0.06\textwidth}p{0.06\textwidth}}
         \toprule
   \textbf{Family (ID)}& \textbf{Parent (ID)}& \textbf{Child (ID)}& \textbf{Race}& \textbf{School} \\
 \midrule
  \textbf{Family 1 (F1)}   & Mother (P1) & Boy 12 (C1) & White & Public \\
  \hline
  \textbf{Family 2 (F2)}   & Mother (P2) & Boy 10 (C2) & White & Public \\
  \hline
  \textbf{Family 3 (F2)}   & Mother (P2) & Girl 11 (C3) & White & Charter \\
  \hline
  \textbf{Family 4 (F4)}   & Father (P4) & Boy 12 (C4) & White & Public \\
  \hline
  \textbf{Family 5 (F5)}   & Father (P5) & Boy 10 (C5) & White & Private \\
  \hline
  \textbf{Family 6 (F6)}   & Mother (P6) & Girl 10 (C6) & White & Public \\
  \hline
  \textbf{Family 7 (F7)}   & Mother (P7) & Boy 11 (C7) & Asian & Public \\
    \hline
  \textbf{Family 8 (F8)}   & Mother (P8) & Girl 10 (C8) & White & Public \\
  \hline
  \textbf{Family 9 (F9)}   & Mother (P9) & Girl 10 (C9) & Asian & Private \\
  \hline
  \textbf{Family 10 (F10)}   & Mother (P10) & Girl 10 (C10) & White & Private \\
 \bottomrule
    \end{tabular}
\end{table*}

\section{Method}
We conducted a participatory design study with children and parents to understand students' needs and preferences from a homework companion robot supporting their homework experiences. The study was conducted on a collaborative whiteboard application and online video calls. Ten children participated in our study and attended a two-hour design session. During the sessions, children shared their current ways of homework and school routine, discussed how they receive homework support from their teachers or family members, and participated in co-design activities to discuss interaction features for an augmented homework support tool to design the speech and expressiveness of a homework companion robot. The activities were supported by articles and comments prepared by middle-school science teachers. These comments were designed with the purpose to support the social and intellectual learning experience of students. Finally, children demonstrated their final design to their parents, allowing them to discuss the design rationale, get feedback, and make changes. We conducted a qualitative analysis of the interviews and design outputs to identify needs that inform design recommendations for a homework companion robot that supports students homework experiences.

\subsection{Participants}
We recruited ten families (See Table \ref{dem-table}) with children aged 10--12 ($M = 10.6$, $SD=0.8$; $Female = 5$, $Male = 5$) from Midwest United States, through teacher contacts at local schools, community centers, newsletters and university employee mailing lists. For a family-centered design approach, we encouraged the participation of both children and their parents. All children in the study participated with one parent ($Mother = 8$, $Father = 2$).
Notably, C2 and C3 were siblings that attended separate sessions as they were able to provide diverse perspectives due to their age difference and attendance to different schools. Their mother, P2, accompanied both C2 and C3's sessions. The siblings did not observe each others sessions. 
All participants were students from different schools (ranging from private schools, charter schools, public schools with an array of different focuses including liberal arts, project based education, religion affiliated, gifted and talented students, or Montessori education) allowing to capture a diverse educational background.

\subsection{Procedure}
Each family attended one online session lasting two hours, which included five activities: (1) context setting interview (10-mins), (2) students' current ways of homework support (10-mins) (3) co-design homework support comments and explore teacher generated comments (30-mins), (4) design the robot's speech and expressiveness in a co-design activity (40-mins), (5) presentation to parents and parent interview (20-mins). 
%
Children individually attended the first four activities, and parents joined the fifth activity. The robot joined the video call in Activity 4 and its speech and movements were controlled by a trained Wizard of Oz (WoZ) operator. The first author facilitated all the design sessions, and one of two trained study team members supported the WoZ actions. The sessions were audio-video recorded.

\paragraph{Robot and Study Resources}
All resources referred in this study are shared for open access~\cite{Cagiltay_osf}\footnote{Open Science Repository for study resources: \url{https://osf.io/8ua7m/?view_only=99b18fa995c3497891a7e953d5075593}}. 
We used a collaborative whiteboard application Miro~\cite{miro} and a Zoom video call~\cite{zoom} to conduct the design sessions. The online whiteboard included a section block for each activity, and the sections were revealed by the facilitator (See Fig \ref{fig:interfaces}). We used Misty II~\cite{Misty} as a robot platform and generated its voice using Google Cloud Text-To-Speech, Wavenet voices~\cite{TTS}. 

\paragraph{Wizard of Oz Control} The social robot was controlled by a Wizard of Oz (WoZ) operator through a dedicated control interface (See Fig \ref{fig:interfaces}). The operator had access to the collaborative whiteboard (Miro), and followed a protocol for generating and executing the comments. After the students finalized writing their comments in Activity 3, the WoZ operator copied these newly written comments and prepared them to be displayed in Activity 4. In Activity 4, if the student made any changes to the comments, the WoZ operator was able to monitor these changes through the Zoom call and Miro board and implement them to the WoZ interface. The updated comments were quickly prepared to be expressed on the physical robot, allowing for real-time feedback and an iterative design.


\subsubsection{Activity 1: ``Tell us about how you do homework!''}
We conducted brief interviews with children to gain context of their current school routine and types of homework they get assigned to. The facilitator asked questions to the child and noted the answers on sticky notes on the collaborative whiteboard. Some example questions related to assigned homework were \textit{``How does a typical school day look for you? Tell me about your daily homework from your classes. What formats do you get these assignments?''} We also followed up with questions related to their homework habits, such as \textit{``Where do you typically do your homework? What type of feedback for your homework works best for you?''}
\subsubsection{Activity 2: ``Let's look into how someone can guide you...''}
The goal of this activity was to familiarize the child to using the collaborative whiteboard, gain a brief understanding of their current ways of homework guidance, and act as a transition activity before starting the co-design sessions in activity 3 and 4. In this activity, the child was presented with a short article, assigned by a middle school science teacher (from ~\cite{pathwise}). The article was related to a grade appropriate science topic, titled ``Fossils.'' We presented page one of the two page article and asked the student to read the article and underline the sections they \textit{``find confusing, need more clarity, have a question about, wish there was more explanation, or didn't know the meaning.''} After reading the article, the facilitator asked the student to summarize their understanding and to describe the reason for underlying specific sections. The facilitator then asked follow up questions about their current ways of support, such as who the student would ask guidance from to resolve their issue (e.g., teacher or family members) or what type of support would work best for them.

\begin{figure*}[t]
    \centering
    \includegraphics[width=\linewidth]{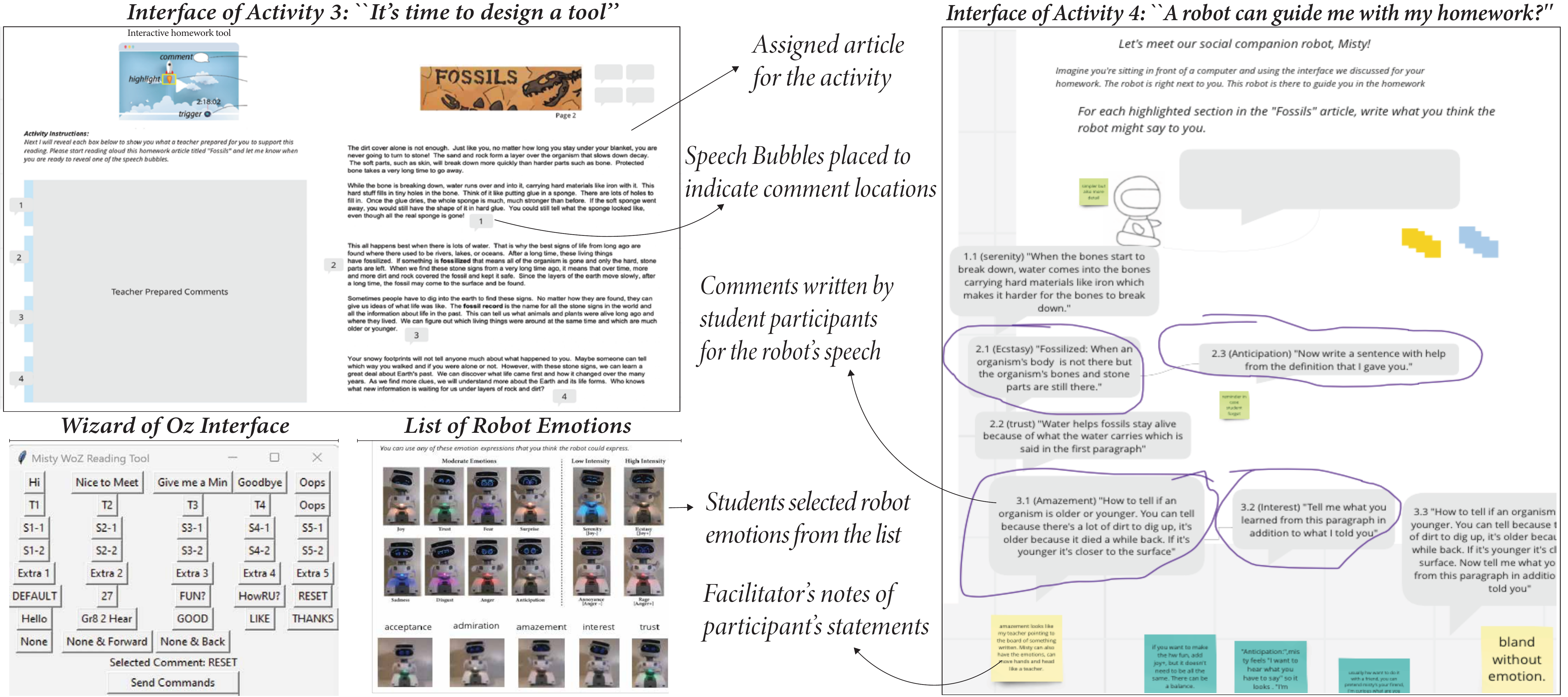}
    \caption{\textit{(Activities 3 and 4)} Articles assigned to students on the Miro Board with speech bubbles, the Wizard of Oz Interface, and a list of robot emotions that were presented to the students, robot's speech and expressions designed by students.}
    \Description{This figure has four main panels. Panel 1 shows a whiteboard interface illustrating the reading tasks that were used in the study, annotated with the text "assigned article for the activity" and "speech bubbles placed to indicate comment locations. Panel 2 illustrates a screen shot from the Wizard of Oz interface. Panel 3 illustrates a list of robot emotions, and annotated with the text "students selected robot emotions from the list." Panel 4 illustrates activity 4 on the whiteboard application, with relevant comments created by the children and post-it notes of the facilitator. This panel is annotated by the text "comments written by student participants for the robot's speech" and "facilitator's notes of participant's statements."}
    \label{fig:interfaces}
\end{figure*}

\subsubsection{Activity 3: ``It's time to design a tool!''}
The goal of this activity was to design interactive features for a homework support interface, where teachers would assign students homework augmented with supportive questions and comments. In this activity, students were presented with page two of the article titled `Fossils,' with four annotated speech bubbles placed on paragraphs of the article (See Fig \ref{fig:interfaces}).
The activity task was described as the following: \textit{``We are going to design an interactive homework guidance tool together. Imagine your teacher assigned you this homework. You log on to an interface where your teacher highlighted some sections in the article, and left speech bubbles.''} Later, the students were tasked to read the article until they reach a speech bubble, and on a sticky note  write down \textit{``what comment or question your teacher might have left in this speech bubble to support your homework experience.''}

After students finished writing their comments for all four speech bubbles, the facilitator revealed a pre-written teacher comment for each one. The facilitator first asked the student's impressions on the teacher-prepared comment, e.g., whether it was relevant, useful, or helpful. If preferred, students had the opportunity to update their own comments and write a new version. This activity was repeated until all four teacher comments were revealed and the student had an opportunity to revise their own created comments. 
Notably, the teacher generated comments included in this study were co-designed by teachers participating in prior work~\cite{pathwise}. Middle school science teachers authored comments for articles and class assignments of their own choosing, tested them on the social robot, and refined them to finalize the comments. Thus, the comments presented to students were carefully designed by teachers with the social robot's interaction in mind.
The facilitator then asked follow up questions for homework tool features, for example \textit{``How do you imagine interacting with the speech bubbles? What type of features would you want this homework tool to have?''}

\subsubsection{Activity 4: ``A robot can guide me with my homework?''}
Students co-designed the speech and expressiveness of a learning companion robot that would support their homework experience. Students used their own comments written in Activity 3 to design the speech and expressiveness of the robot in Activity 4. This co-design activity iterated between three main parts: 
\textbf{(1)} Listen to the \textit{teacher} generated comments from the robot Misty, 
\textbf{(2)} Listen to the \textit{student} generated comments from the robot Misty, 
\textbf{(3)} Make any changes to the student or teacher generated comments.

At the beginning of Activity 4, facilitators informed children that they will now be designing interactions for Misty, a social homework companion robot that will be joining the call. Once the robot joined the Zoom call, the robot greeted the child, asked their name, and waited for a response from the child. The facilitator then described the robot's capabilities, by saying \textit{``Misty can speak, move its arms and body, show facial expressions, and listen to you''} and described that the robot is \textit{``training to become a homework support assistant''} for students their age and \textit{``we need your help to train the robot and you're the expert.''} 
We then followed up with questions capturing the child's impressions of the robot, e.g., \textit{``Imagine you're sitting in front of a computer and using the homework tool we discussed. The robot is right next to you to guide you in the homework. (1) Would you want a robot that would support your homework? (2) Where do you imagine interacting with this robot? (3) How would you want the robot to guide you?''}

The collaborative whiteboard for Activity 4 displayed a list of 16 static figures of robot emotions available for reference. These emotions were designed for the Misty robot adapted by the Plutchik's emotion wheel model~\cite{plutchik1980general}, validated and used in ~\cite{zhao2020emotion, emotion-idc}. Students were able to review the emotion list and pick an emotion for the robot to display while saying their comments. The WoZ operator would then execute the expressive speech (comment with emotion) on the robot in real-time, so the child could observe the robot's dynamic color change, facial expressions, body movements, gestures and head orientations while hearing the comment. This demonstration allowed for students experience robot behaviors around the sample reading and try any changes they felt necessary (See Fig \ref{fig:interfaces}). 
In short, children iteratively tested and edited their comments and added emotions, listened to it from the robot, and repeated these actions as many times as they preferred. 
After completing the co-design activity, the facilitator asked wrap-up questions to capture the child's impressions of the robot, e.g., \textit{``Now that you have spent time with the robot and seen how Misty could interact with you, what do you think about it? What other homework related or non-homework related features would you want this tool to have? What other ways would you want the robot to guide you?''}


\subsubsection{Activity 5: Demonstration to Parents}
To follow a family-centered design approach, we invited parents to participate in a part of the co-design study, allowing their contribution to their child's design and share any suggestions, concerns or improvements. After the student completed their designs in Activity 4, the facilitator informed them that they will be presenting the design to their parents and asked them to select the comments they would like to showcase. Once a parent joined, the student presented and described the comments and expressions they designed for the robot. 
Afterwards, the expressions and comments created by children or teachers were executed on the robot. The facilitator conducted semi-structured interviews while the parent observed the robot's comments and the child and parent discussed robot design improvements. Parents were asked their thoughts, recommendations, and concerns about the robot's speech, expressiveness, purpose, personality, etc. Parents could make adjustments to the comments if preferred. 

Finally, we conducted brief interviews with parents to better understand their perceptions, preferences, or concerns towards learning companion robots for their children, e.g., \textit{``What type of support works best for your child? How do you think a social robot could deliver that support? Do you have any concerns for having a robot in your home?''} 



\subsection{Qualitative Analysis} We conducted a reflexive \textit{Thematic Analysis} following the guidelines presented by \citet{braun2019thematic} and \citet{mcdonald2019reliability} on the interview transcripts and design outputs from Activities 3, 4 and 5. We only included descriptive information from Activity 1 for demographic context and we did not analyze Activity 2 since it was intended as a warm-up activity. The first and last authors were familiarized with the data by facilitating the design sessions, transcribing, and reviewing the session recordings. The first author initially generated potential codes on a digital whiteboard and discussed candidate themes with the study team, later revised and combined related themes. Finally, the authors reviewed, defined, and reported the themes as findings. When reporting our findings, we use participant ID's (See Table \ref{dem-table}) where children are referred to as C1-C10, parents as P1-P10 and families are referred to as F1-F10.

\section{Results}
We report the results from our co-design sessions categorized in two themes. First, we report how children associated specific robot emotion expressions with specific types of homework support phrases and how the emotions can impact the motivation and affective state of students. Second, we found that children either perceived the robot as a \textit{dedicated homework assistant} that would not only motivate them to do their homework, but also provide companionship outside of homework activities through meaningful social interactions, or as a \textit{shared learning resource} for students in the classroom. 
We elaborate on each theme with quotes and observations from the co-design activities. 

\subsection{Theme 1: Robot Comments and Emotions as Central to Supporting Learning Resources}
In Activity 3 and 4, children created their own homework support comments, then added any emotional displays they would like and could see the robot enact each comment/emotion. We found that children created comments to \textit{summarize}, \textit{define}, or \textit{expand on} content in the reading, and that \textit{ask questions} or \textit{relate} the content to children's lives. Children also saw strong similarity in the comments they wrote to the sample teacher comments and found adding the right emotional displays as crucial to properly delivering comments.

\subsubsection{Children created comments similar to teacher's comments} 
The first design activity, Activity 3, was tailored to explore the preferences of children for the types of comments they would like to hear to help them with homework. 
During this activity, children wrote various types of comments for the speech bubbles. More than half of the children created a comment that made a simple summary (C3, C4, C5, C6, C7, C9, C10), a comment that relates the content to the student's life (C1, C3, C5, C6, C8, C9, C10), or a word definition for students to understand (C1, C4, C5, C7, C9, C10). Children also created comments that include prompts for the student to give an example (C1, C6, C9, C10), state a fun fact (C3, C5, C8, C9), questions to excite the student on the topic (C3, C6, C10), analogies to illustrate concepts (C1, C8), or share extra information not captured in the text (C6, C7, C10).

After creating their own comments, children read sample teacher comments for the reading, and more than half of the children (6/10) felt their comments were similar to the provided sample teacher comments, derived from prior work with teachers (C1, C2, C4, C6, C9, C10). For example, C6 said that \textit{``it sounds like something my teacher would say''} and C1 told us, \textit{``that's definitely a question my teacher would ask.''} Children found that teacher generated and student generated comments shared many similarities in their language, goal, and sentiment (e.g., Table \ref{student-teacher-comments}). Some children (C3, C5, C8) immediately noticed these similarities. For example, after the first teacher comment, C3 stated, \textit{``that's a good one! I feel like I had something similar for my second one.''} She compared her comment and the teachers by saying \textit{``it (teacher's comment) is probably better than mine. It is way shorter than what I was trying to get at, except much better phrased than mine.''} Some children expressed that they \textit{``like how [the teacher comment] challenges''} the student (C6, C8) and makes them \textit{``think about the topic''} (C1, C6, C10) but it is also \textit{``age-appropriate''} (C3). Children interpreted the purpose of the teacher comments as helping students focus on the homework (C1, C9, C10), to bolster their interest in learning about the topic (C4, C7, C8), and acting as a reminder and summary of the assignment (C1, C3, C7, C8, C10). For example C1 told us the comments felt like a teacher is asking questions \textit{``just to check that you've been paying attention to the article.''} Similarly, C9 said \textit{``it's pretty good because it makes sure that you're paying attention, and you're following along.''} C4 told us they felt the comments were \textit{``questions for you not to get graded on''} but designed to help students \textit{``get curious''} about the topic. C10 said that it's \textit{``kind of cool that the teacher can leave comments, then it will kind of get me thinking about what I learned.''} In short, students saw robot comments serving a purpose of both helping them understand the content but also to help drive their interest and engagement with the subject, and students seemed to find a similar style of comment writing between their own comments and the sample teacher comments. 

\begin{table*}
\centering
\caption{Comments that were prepared by teachers and used in the study. Selected examples of comments prepared by students.}
\small
\begin{tabular}
{p{0.025\linewidth}p{0.71\linewidth}p{0.19\linewidth}}        
\multicolumn{1}{c}{} & \textbf{\textbf{Teacher Prepared Comments}} & \textbf{\textbf{Comment Type}} \\ 
\hline
TC\#1 & Can you think of other living things that might leave a fossil behind? Remember, it takes a long time. & Question to reinforce learning \\ 
\hline
TC\#2 & A new vocabulary word! Make a vocabulary card with the word definition and use it in a sentence of your own. & Word definition \\ 
\hline
TC\#3 & Now, tell me three things you learned about fossils. And tell me one thing you are still wondering about. & Question to reinforce learning \\ 
\hline
TC\#4 & That's great! And we will be looking at real fossils tomorrow. Remember, for something to be fossilized, it has to have been protected, usually by sediment or water, for a very long time. See you tomorrow- be ready to get dirty! & Fun fact related to classroom activities \\ 
\hline
\\& \textbf{Selected Examples from Student Prepared Comments} \textit{[Child ID]} & \textbf{Comment Type}\\ 
\hline
SC\#1 & An organism is a single celled life form, a plant, or an animal. \textit{[C9]}  & Word definition\\ 
\hline
SC\#2 & Think of fossils as the layers of a notebook. You have the hardcover so the pages aren't harmed by water or heat, and we have the center that's soft. So you have to protect it with a harder exterior.    \textit{[C8]} & Analogy to reinforce learning        \\ 
\hline
SC\#3 &What kind of fossils would you want to see and learn about? What is your favorite kind of dinosaur? Have you seen any fossils of your favorite dinosaur? \textit{[C3]}  & Question relating to student's life        \\ 
\hline
SC\#4 & People who dig in to the earth to find these signs are called paleontologists. \textit{[C10]}& Fun fact  \\
\hline
\label{student-teacher-comments}
\end{tabular}
\end{table*}

\subsubsection{Children found robot's emotions to be important when providing homework support}
In the second design activity, Activity 4, children listened to the robot ``speak'' the teacher comments and their own comments. Children's first reactions to hearing comments from the robot were mostly focused on the clarity of speech and their ability to understand the robot. C10 said that \textit{``the message is delivered as it was intended''} and C4 expressed that \textit{``it was very clear, I could hear it well, and [Misty] said it perfectly.''}  After establishing the clarity of verbalization of comments, children (C1, C3, C4, C6, C7) strongly emphasized adding non-verbal or emotional features of the robot, including facial expressions, head, arm and body movements, or changes in the chest light, to accompany each comment. Half of the participants (C6, C7, C8, C9, C10) expressed that \textit{``Misty has human-like attributes,''}(C7) \textit{``almost looks like a real person,''}(C10) and \textit{``does not look like a robot... That makes me feel more comfortable.''} (C7).
Two children (C5, C10) associated the human-like features with the robot's head movements, as C5 told us, \textit{``(Misty) moved her head in a way that she was like stating something to the class.''}
%
Two children (C6, C9) associated the human-like features with the robot's voice. For example, C6 said, \textit{``she has an expression like my teacher. Sounds kind and stern. It's very engaging.''} Many children (C1, C3, C5, C7, C10) shared that verbal elements such as the speech expressiveness, pitch, and intonation could contribute to a positive homework experience.
%

During this activity students selected specific robot emotions to accompany the comments they had written. Out of 74 comments created by 10 children, 47 of the comments were tagged with an emotion and 27 comments were left with a neutral expression. Most commonly used emotion expressions for comments were: interest (9), trust (8), ecstasy (7), serenity (5), joy (4), admiration (4), surprise (4), acceptance (3), amazement (2), anticipation (1). Two children (C4, C7) preferred to use multiple emotions for different parts of a comment, allowing the robot to shift between emotions throughout its speech. 
 For some children, seeing accompanying emotions from the robot instead of neutral expressions may be important for understanding the robot, as C5 told us, \textit{``it's a lot easier to understand the words when it has facial expressions. Otherwise it's kind of just like saying words out of the speakers and just having normal eyes just waiting there.''} C4 said, he would think harder and give more details in his response to the robot because the emotional display makes the robot appear more interested in what he has to say.
Overall, children associated many of the emotional expressions with specific types of homework support. Table \ref{table_emotions} shows example quotes from children related to the associated emotions with different types of homework support. These include children who associated the `interest' emotion with question asking (C4, C5, C8), the `surprise' emotion with learning a new word (C1), the `ecstasy' emotion with fun and interesting homework content (C7), the `trust' emotion with stating a fact (C9), and the `amazement' emotion with fun facts (C9) or excitement about the topic (C7). However, children also expressed that having the same emotion might get repetitive and boring over time, and suggested substituting alternative emotions when needed.




\begin{table*}[t]
    \caption{\textit{Children associated certain robot expressions with different types of homework support comments.}}
    \label{table_emotions}
    \centering
    \small
    \begin{tabular}{p{0.08\linewidth}p{0.11\linewidth}p{0.75\linewidth}}
         \toprule
   \textbf{Expression}& \textbf{Comment Type}& \textbf{ Supporting Quote} \textit{[Child ID]} \\
         \midrule
 Interest & Question & ``When you \textbf{ask a question} it would look like \textbf{interest}. Misty looks interested or curious about your opinion... Having all questions with interest might get boring and repetitive, so sometimes it can be happier emotions like, joy, serenity, surprise.'' \textit{[C4]}\\ \hline

 Surprise & Definitions & ``I think I would like the \textbf{surprise} one for all the \textbf{word definitions}. I think it'll be pretty funny. Because Misty would be like `Wow! You learned a new word!' '' \textit{[C1]} \\\hline

 Ecstasy & Fun & ``If you want to make your homework like more interesting and more \textbf{fun}, then you can add joy plus (\textbf{ecstasy}). But like, it doesn't always have to be that fun, you can choose which one you want.'' \textit{[C7]} \\ \hline
 
  Trust & Facts &  ``I think, for the ones that I wrote, they were just saying a \textbf{fact}, so maybe \textbf{trust} would be the best for them, because trust is like the most normal, like, does not have a giant smile. I think all the facts should probably be trust.'' \textit{[C9]}\\

 \bottomrule
    \end{tabular}
\end{table*}

We also found the emphasis on the application of emotional expressions was impactful when there were incongruities between the robot's emotion expressions and verbal comments. To some children the written comments from the robot sometimes did not match the emotional expression that children expected. C1 described this as, \textit{``it's different to hear the written comment from the robot, because [Misty] looks more joyful and interested in the topic.''} Many children (4/10) expressed that the comment and the emotion expressions were typically congruent with each other, e.g., (C4) \textit{``the emotion matches what she's saying''} and C8 suggested that congruency of expressions might be perceived as an indicator of the child's performance, i.e., \textit{``when the emotions and the visual effects match, like the face and lights, this might give an effect, good or bad, related to your performance. If she was green, it could make you happy.''} When emotion expressions did not match verbal comments, children thought the robot was bored (C10) or not interested (C9). 
While other non-verbal cues were available, children seemed to find facial expressions sufficient to properly match emotions to the intonation of the verbal comment (C1).
To mitigate incongruities in the emotion and speech, children experimented with different facial expressions. For example C8 wrote the comment \textit{``Find a type of track of an animal. You can see if it was with any other animals. You can figure out the size of it and where it was going.''} and added the emotion ``trust.'' However, after seeing the robot enact this comment with the emotion, C8 said, \textit{``I think I might change the emotion just because I don't think the expression quite matches what it's saying... It might have been confusing to some people about the topic if the facial expression was like that for that question.''} To fix this mismatch, C8 instead chose the ``serenity'' emotion expression and said they ``like it'' after re-playing the comment with the expression. In this way, we see that some children emphasized how the emotional behaviors were important to help them understand the robot. 
\subsubsection{Children felt robot's emotions could deliver motivational and affective support}



Many children (C1, C6, C7, C8, C9, C10) suggested that the emotional expressions might support their motivation towards homework. Some specific emotional expressions for the robot, such as `interest' and `anticipation', were associated with the perception of the robot being curious about the reading (C7) or interested in the child's opinions and what they have to say (C4, C5). In contrast, the default `neutral' expression was associated with ignoring the person (C9) or being bored (C10). Some children suggested that appropriate emotional expressions inferred the robot was \textit{``being excited with somebody''}(C1) and expressions made it \textit{``fun and enjoyable to do the homework''} (C6). The excitement expressed by the robot would \textit{``help to get started on my homework''} (C6) or help the student to be \textit{``motivated to complete the homework faster''} (C1).
C7 shared that the robot's head movements and facial expressions \textit{``create more joy''} and can motivate the child to \textit{``finish up''} when homework is `boring.' C7 thought the neutral expression might, \textit{``focus you up,''} to help the child persist on assignments, where the robot might be perceived as \textit{``pretty much like a teacher at home.'' }
Two children expected to hear motivational phrases from the robot while doing homework. C1 said, \textit{``if the robot can read your emotions and tell if a student is sad or confused, it could motivate the student.''} Similarly, C7 said homework could be more joyful if the robot \textit{``comes up to you and tell you-- you're doing great!''}
C6 shared that the robot's `smile' made her \textit{``feel like I did something right.''} She stated that this feeling would \textit{``help me keep going, and like just a little bit of encouragement.''} C6 followed up with a scenario where she might have \textit{``got a problem wrong''}, in that case Misty would say `that's OK', and give a `soft smile.' C6 expressed that this smile \textit{``would make me feel better, like, even though I got one wrong. It's okay. I can just put in the right answer and move on.''} C4 further discussed that the robot's lack of engagement or neutral expressions could negatively impact motivation to work on their homework, and told us, \textit{``if Misty doesn't really want to do this, it's not going to be fun for me, because Misty is explaining and asking these questions for me. If misty is sad, or mad, then does it mean that Misty doesn't want me? Doesn't want like to do these questions and doesn't want to be here right now?''} C6's parent, P6, similarly commented on the neutral expressions affecting her mental state, i.e., \textit{``neutral face would make me feel like I was doing more work, and that isn't that fun. It would kind of bring down the mood if it was just like straight face. I really like the smile.''} Overall, children associated the robot's facial expressions, head movement, motivational speech and phrases as ways of supporting students to be motivated for homework.


These motivational supports also seemed to extend beyond simply helping with homework. Five children (C3, C4, C6, C7, C10) expressed that to some extent, the robot's emotion expressions could impact the student's affective state. C3 stated that having the robot \textit{``would relieve some of the stress and anxiety I have over not finishing things on time.''} C7 expressed that the robot's facial expressions gave them `relaxation' and `comfort' in cases where \textit{``you have a very bad day at school. Misty could like cheer you up with that facial expressions, and her voice.''}
C10 and C4 both discussed how a happy robot could make a child happy and when the robot is `excited' then the student will be excited as well. C4 discussed this issue by contrasting negative, neutral, and positive emotions, \textit{``if it's like neutral and boring, then you don't do your best work. So if there's expression, then it makes you like happier and then very happy. You do better in your homework.'' }
If the robot shows negative emotions including \textit{``scared, or something, like sadness, disgust or anger, I would be like, `I don't want to do this (homework). It's not fun.' But if Misty does something like happy, then it makes other people understand that Misty is happy, and then you're happy.''}

\subsubsection{Parental perspectives for the robot's expressions}
Most parents (6/10) enjoyed the comments and expressions from the robot and said it's \textit{``a fun and engaging layer to provide more information''} (P6). However, one parent, P2, expressed skepticism towards the robot's emotions after observing the robot's interaction for the first time, saying that \textit{``well as an adult, I'm not the right person to wonder about robot emotions, because I don't really think of robots so much as having emotions. But [C2] is a child and I think they would notice that[emotions] more than I do.''} P2 later expressed that they \textit{``got used to it over time.''} Some parents (P2, P3, P6, P7) shared that the robot's emotions could be used to motivate their children as an \textit{``enjoyable activity for a child to pay more attention and be more willing to answer a question from a robot than just reading it on a sheet of paper''} (P2). P6 stated that their child typically does not need support during their homework, but needs motivation to get started. The robot could use \textit{``encouraging speech''} (P3) to motivate the child \textit{``to start their homework''} (P6) which could also make the homework more \textit{``exciting by breaking the monotonity''} (P7).

\subsection{Theme 2: Robot as a Dedicated Homework Assistant or Shared Learning Resource} 
Children's current homework habits aligned with their preferences for homework assistance from the robot.
Participating children were grouped in three categories based on their homework habits: children that (1) don't get homework assigned and complete assignments at school (C10), (2) sometimes bring incomplete or optional schoolwork home to complete them (C2, C3, C8), or (3) have regular homework assigned to them (C1, C4, C5, C6, C7, C9). This variety captured diverse preferences towards the robot's role as a homework assistant.
 Three children (C2, C9, C10) expected the robot to be a shared resource at school, and the remaining (7/10) expected the robot to be a dedicated assistant for themselves at home. For the in-home homework assistance, children expected the robot to provide companionship during and after homework activities through meaningful social interactions.

\subsubsection{Children imagined the robot as a dedicated homework assistant and friend for the child at home}
Most children wanted to have a dedicated homework companion for themselves at home and form friendships with the robot that would go beyond homework support. Children mentioned that teachers or parents sometimes might be busy and won't have the time to support the student at that very moment, which was one of the key motivations for children to have a \textit{``dedicated homework buddy,''} e.g., C1: \textit{``somebody that's always like ready for doing homework, and doesn't need to do anything else. My mom, she has like work so she can't always help me with homework. That's not her job. So I think the robot will definitely help.''}
C3 expressed preferences towards having the robot personalized to themselves, \textit{``I would like it as a friend. That is your personal friend, that you don't have to share with anyone else that it's like programmed for you.''} C3 followed with, \textit{``if the robot was just like school property I would be fine with it. But I feel like I would rather have it more of like a personal thing than something everybody is using, because then I would develop a more personal connection with it. And that's something that is important to me.''}
Three children shared that they would prefer the robot to be by their side for companionship, e.g., \textit{``when I'm doing homework at home, and there's no one around, and my teacher's not here and my classmates aren't here''} (C6). Similarly, C7 described that \textit{``usually when you do your homework, you kind of want to do it with a friend. You could just pretend like she's your friend and you can just tell her[program Misty] to use `anticipation' and ask `What are you thinking of?' Make them[students] feel like they have a forever friend next to them helping them.}'' C4 similarly described the joy of having a conversation companion with them while doing homework, \textit{``you get to talk to someone or someone gets to tell you stuff instead of you just doing everything by yourself. That gets boring, it's not as fun. It's more fun to talk to someone than just sitting there doing it all by yourself.''}

\subsubsection{Children imagined the robot as a shared learning resource for students in the classroom}
Children that preferred the robot to support them at school envisioned it as a shared learning resource for peers to interact with in the classroom ``because we don't really get assignments that we have to take home'' (C10).
Children saw the robot as a supportive assistant for the teacher as well, i.e., someone to go to when the teacher is busy, C2: \textit{``if I had a question but my teacher was busy, the robot can help me understand what my question was and answer my question.''} Similarly, C10 described \textit{``sometimes our teachers are pretty busy, and we don't want to bother them if we don't have to. So you kind of figure it out on your own, either with your friends or maybe with Misty at school.''}
Children also envisioned interacting with the robot in groups. C2 wanted the robot to be \textit{``at school, at my desk with a group of kids that will be with me around my desk.''} C10 imagined the robot to be in groups of 6 -- 12 students, and anything more than that could get \textit{``a little bit chaotic.''} In these group settings, the robot would be \textit{``directed at one person, but if there's someone else who's working on the same thing then Misty can also help them.''} C10 emphasized that group members would need to take turns where \textit{``everyone can get to interact with Misty a little bit on their own turn, in their own assigned time.''} For shared interactions C10 discussed that larger groups should \textit{``understand that using Misty is like a privilege, and they are not entitled to Misty and you have to share.''}

Motivations of C9 for having the robot at school was different than C1 and C10, and mainly related to trust, \textit{``you don't know if you can trust them[Misty]. You don't know their personality.''} Even though C9 expressed positive feedback in the design sessions for the robot's emotion, speech, and use case as a homework companion robot, C9 was skeptical about having a robot in their home. She described it in the following way \textit{``the robot would be located at school, because I would be scared to be alone with robot. I imagine I would probably be sitting in my classroom because for some reason I'd be pretty scared to be in the room (at home) alone with robot.''} P9, the parent of C9, however, was supportive of having the robot at home and did not voice any concerns as a parent. C9 described that she would need time to get to know the robot by \textit{``talking normally for about half an hour''} and having \textit{``ice breaker activities''} could help her to gain trust.

\subsubsection{Parental perspectives for in-home homework companion robots}
When asked about their ways of homework support, children explained that they typically seek support from parents or older siblings at home (7/10), reach out to their teachers over email (1/10), or ask help from teachers at school (3/10). Parents were supportive of the idea of having the robot either at their home or at the child's school and placed in rooms their child typically does their homework, e.g., in a study room (P9), living room (P7), or bedroom (P6). P2 told us it could \textit{``relieve stress from parents at home''} and explained \textit{``you know you're trying to get dinner made and get the load of laundry in, and all these things, and but still want your child to be working on their homework and finishing things.''} P2 further recommended it could alternatively be used in an after school program located at \textit{``community centers where kids might not have their parent or teacher around to guide them through things,''} \textit{``instead of one robot per child.''} One parent (P6) stated that their family struggles with screen time and try to reduce the time they \textit{``interact with electronics in general.''} However, P6 perceived the robot as a \textit{``different type of technology. An educational robot would not be a bad thing.''}

Two parents (P2, P5) expressed concerns about robot `abuse' (P5) or how to respond if children \textit{``mess with the robot''} (P2). Parents preferred to do \textit{``minimal maintenance on the robot''} (P5), but have instructions on how to support the child in case there are technical problems.
Three parents (P2, P5, P6) mentioned privacy and video recording concerns and how it \textit{``it might be creepy if it is recording a video of the child while they're working''} (P2), but P5 also mentioned that it isn't that different than other technology they use daily. Three parents (P5, P6, P8) recommended features that related to adapting and personalizing to the child's habits and education level. P8 summarized this as \textit{``if it's too easy, it's also easier for them to disconnect. If it's too hard, they might get really frustrated. So I think, having it be customized to [C8]'s learning would be really important for it to be successful.''}

\section{Discussion}
We explored children's design preferences for social and intellectual supports while learning with robot-augmented homework experiences. Our research question was, \textit{RQ: What needs do students have for interactions with a socially situated homework companion robot?} We firstly found the robot's emotions were perceived as an important part of the homework experience, in which these expressions can support students' affective state and motivation to do homework. Secondly, children imagined the robot either as a dedicated homework assistant at their home, or as a shared learning resource in the classroom.
We interpret our findings from the perspective of students, supplement them with parental and teacher perspectives, and translate our findings into design guidelines for educational homework companion robots.

\subsection{Design Guidelines for Educational Robots}

\subsubsection{\textbf{Design for Emotionally Expressive Homework Support Comments to Support Learning, Engagement and Motivation}}
We identified that children in our study prepared socially supportive comments that were similar to what the teacher had prepared from prior work~\cite{pathwise}. This was an important validation since none of the children were students of the teacher, children were from different school grades, and many did not share the same education system. This indicates that with minimal practice, teachers can create comments that are received well by students when expressed by a homework companion robot. 
However, children were more interested in adding associated emotions and non-verbal behaviors to these homework comments than they were in creating the verbal content of the comments, and we found most children carefully annotated robot comments with appropriate emotions until they felt satisfied with the emotion-comment pairing.
Our findings indicate that children prioritized \textit{``how'' they want to hear and experience} a comment, rather than \textit{``what'' they want to hear in the comment}, similar to teachers' perspective in~\cite{ceha2022identifying}. 
However, this contrasts with our prior work where teachers rarely added emotional expressions~\cite{pathwise}, and focused more on the homework support comments’ content and child-friendly tone.  
Given that students prepared verbal portions of comments with similar tone as teachers, this points to a need for tailored support for teachers to incorporate robot emotion when preparing socially and intellectually supportive homework comments. 

Similarly, when children talked about factors that could support their homework \textit{motivation}, they more often referred to the robot's emotion expressions instead of the speech content. This aligns closely with the design goals for social robots that support children's learning~\cite{davison2020designing}, i.e., robot's emotional response display, offering emotional support to the student, and engaging in social activities beyond learning.
However, as highlighted in our results, some adults may not be as sensitive to the robot's emotions as children are, but children may interpret the robot's lack of emotion as disinterest or boredom, which may cause them to lose interest and motivation in their homework. 

\textbf{Need:} Children are sensitive to the robot's emotion expressions and need appropriate emotions displayed by the robot to feel supported and motivated in their homework experiences. However, adults and educators might not pay much attention to emotions, and prioritize \textit{what} the robot will say, rather than \textit{how} the robot delivers it (i.e., with added emotion).

\textbf{Recommendation:} Provide structured guidance for teachers to include appropriate \textit{emotional} or non-verbal cues when creating robot assisted learning content. The speech design of a homework companion robot should add equal emphasis to emotional expression to more wholly support learning interactions and motivation.

%

\subsubsection{\textbf{Homework Companion Robot's Role as a Dedicated Assistant at Home or Shared Learning Resource at School}} Our findings highlighted contrasts between what children expect from a homework companion robot's role in formal learning interactions. 
Some children envisioned the robot as a shared resource in the classroom, while most children expressed a desire to have the robot dedicated to them and not as a shared entity at their home. This partially contrasts with prior work, where in informal learning settings at home, \textit{e.g., social reading}, children preferred to share the robot with siblings or other family members~\cite{Cagiltay_engagement_2022, offscript_HRI}.

While families in our study generally saw value in having this tool for dedicated homework support at home, there is an important need to integrate these technologies into family life in ways that are supportive of the existing human relationships, and does not get in the way of connection-making between family members. Participating children in our study indicated that they currently receive some level of homework support from their parents or siblings. However, parents and children also noted a desire to have the robot as a dedicated resource, as parents might find homework support time consuming or conflicting with other obligations, which also aligns with~\cite{van_voorhis_costs_2011}. 
As recommended by \cite{van_voorhis_costs_2011}, teachers can design immersive homework experiences for the robot that would not only be customized for the student, but could also allow for a collaborative and meaningful homework interactions including the child's peers or parents. In line with recommendations from prior work, this could potentially support family-centered homework experiences, resulting in more positive emotions and attitudes towards homework. Social homework companion robots can be useful resources to provide direction to students during homework without burdening the parents so that students can link concepts, find value in the homework, and deal with distractions or drops in motivation~\cite{xu_why_2013}. These interactions can also be designed to afford child-robot dyads, child-parent-robot or child-peer-robot triads to allow for enhanced connection-making opportunities with family members in mind. Such interaction designs can facilitate stronger, less demanding, more entertaining and educational interactions that could center the student, while allowing the family to be included as a whole (e.g., older or younger siblings, parents, grandparents, and peers).

\textbf{Need:} Children may have different desires for sharing or not sharing a robot at their home or with classmates at school. Interactions with the robot need to balance student desires for individual use with supporting inter-family connections and engagement around homework.  

\textbf{Recommendation:} To meet the different needs of children, design robot interactions flexibly so the robot can be a shared supportive resource for group interactions in classrooms, afford dedicated companionship for students during formal in-home learning, and afford family-centered homework experiences that allows for connection making between the child-parent dyad without putting the burden on the parents.

\subsection{Limitations and Future Work}
A number of factors limited our work. First, we held the design sessions remotely through a video call where participants could not directly interact with the physical robot, which might have hindered children's ability to fully experience the robot's embodiment and introduced constraints to the design sessions that might not have been present in-person (e.g., as discussed in ~\cite{lee2021show}). The remote nature of the study, however, enabled us to reach a larger and more diverse audience, such as students living in different cities or attending different schools, which wouldn't have been feasible to do in-person. We acknowledge that the small number of participating families that attended from the same region limits the generalizability of our results for families with different racial, educational, and socioeconomic backgrounds. In the next phase of our research, we intend to recruit a broader sample of students from different cities and conduct in-person co-design sessions, where they will be able to interact with and evaluate the robot firsthand. Second, we acknowledge that our findings may be biased by our research interests, as our research is motivated by the goal to design an augmented homework studio for teachers and students where homework support comments can be expressed by social companion robots.
Third, the robot's interactions were not autonomous, but rather controlled in a Wizard of Oz style. This was necessary to rapidly prototype features and express them on the robot for iterative co-design. Future applications of this work will implement fully-automated robot features. Fourth, participation was limited for only one parent and the child, where extending family and household members were not able to share their perspective. 

Future work will focus on developing a prototype homework studio and robot interactions based on what we have learned from educators, students, and parents. We will continue iterative development and co-design sessions with students and teachers to design (1) teacher- and student-facing homework studio prototypes, (2) methods for the robot to respond to both print and digital homework formats, (3) mechanisms for personalization and feedback, and (4) workflows for teachers, students, and family members to partake in the homework experience and interact with the robot. We will investigate how the robot can be used in school and at home, as well as how it can accommodate individual and group interactions.

\section{Conclusion}
We explored children's preferences for a homework companion robot through an online co-design session with children. Children associated robot's emotions with specific types of support comments, preferred the robot to support their affective state and motivation to do homework, imagined the robot either as a dedicated homework assistant at their home, or as a shared learning resource in the classroom. Parents also shared their preferences towards the robot's emotion expressions and having a homework companion robot at home. Overall, this work serves as an iterative step towards participatory design of a homework companion robot and homework studio interface for students and teachers.

\begin{acks}
This work was supported by National Science Foundation award \#2202802. We would like to thank Sylvia Joseph for her valuable contributions in supporting the Wizard of Oz sessions.
\end{acks}

\section{Selection and Participation of Children}
Children aged 10--12 were recruited through their parents who were contacted through teacher contacts at local schools, community centers, newsletters and University employee mailing lists. Parents filled out a pre-screening survey to report their child's demographics. The main inclusion criteria was having a child aged 10--12 and participants were scheduled on a first-come-first-served basis until gender balance and study quota was reached. To obtain informed consent, researchers provided a description of the study and how participants' confidentiality would be protected in publications (i.e., any identifiable information will be removed or anonymized from the data, including videos, images, or transcribed speech). Detailed descriptions of the consent language is shared in the supplemental study protocol materials shared for open access. The study was initiated only after the parents provided written consent and children provided verbal assent. After completing the study, parents received a \$25 digital gift card. The study was reviewed and approved by the University of Wisconsin-Madison Institutional Review Board (IRB).

\bibliographystyle{ACM-Reference-Format}
\bibliography{manuscript}


\end{document}